\documentclass[useAMS]{mn2e}

\usepackage{graphicx,color,soul}
\usepackage{latexsym}
\usepackage{multirow}
\usepackage{amsmath,amssymb}        
\usepackage[draft=false]{hyperref}

\title[Collisional dark matter density profiles around supermassive black holes]{Collisional dark matter density profiles around supermassive black holes}

\author[F. S. Guzm\'an, F. D. Lora-Clavijo]{F. S. Guzm\'an, F. D. Lora-Clavijo\thanks{E-mail:
guzman@ifm.umich.mx (FSG); fadulora@ifm.umich.mx (FDLC)} \\
	     Instituto de F\'{\i}sica y Matem\'{a}ticas, Universidad
              Michoacana de San Nicol\'as de Hidalgo.\\ Edificio C3, Cd.
              Universitaria, 58040 Morelia, Michoac\'{a}n,
              M\'{e}xico.}

\begin{document}


\date{\today}

\pagerange{\pageref{firstpage}--\pageref{lastpage}} \pubyear{2011}

\maketitle

\label{firstpage}


\begin{abstract}
We solve the spherically symmetric time dependent relativistic Euler equations on a Schwarzschild background space-time for a perfect fluid, where the perfect fluid models the dark matter and the space-time background is that of a non-rotating supermassive black hole. We consider the fluid obeys an ideal gas equation of state as a simple model of dark matter with pressure. Assuming out of equilibrium initial conditions we search for late-time attractor type of solutions, which we found to show a constant accretion rate for the non-zero pressure case, that is, the pressure itself suffices to produce stationary accretion regimes. We then analyze the resulting density profile of such late-time solutions with the function $A/r^{\kappa}$. For different values of the adiabatic index we find different slopes of the density profile, and we study such profile in two regions: a region one near the black hole, located from the horizon up to 50$M$ and a region two from  $\sim 800M$ up to $\sim 1500M$, which for a black hole of $10^{9}M_{\odot}$ corresponds to $\sim 0.1$pc. The profile depends on the adiabatic index or equivalently on the pressure of the fluid and our findings are as follows: in the near region the density profile shows values of $\kappa <1.5$ and in the limit of the pressure-less case $\kappa \rightarrow 1.5$; on the other hand, in region two, the value of $\kappa<0.3$ in all the cases we studied. If these results are to be applied to the dark matter problem, the conclusion is that, in the limit of pressure-less gas the density profile is cuspy only near the black hole and approaches a non-cuspy profile at bigger scales within 1pc. These results show on the one hand that pressure suffices to provide flat density profiles of dark matter and on the other hand show that the presence of a central black hole does not distort the density profile of dark matter at scales of 0.1pc.
\end{abstract}





\section{Introduction}

One of the most important issues related to the dark matter problem is that of the dark matter density profiles in galaxies. Most of the analyses involve the study of profiles based on simulations of structure formation and collapse of cold dark matter (CDM). Among the most studied models there are for instance the Navarro Frenk White (NFW) density profile \cite{NFW1996,NFW1997} and Moore's model \cite{Moore}. Within the study of density profiles it is particularly interesting to understand the distribution of matter in the innermost region of the halo, where different models provide different density slopes of the type $\sim 1/r^{\kappa}$; in particular the NFW and Moore profiles show different behaviors $\kappa=1$ and $\kappa=1.5$ respectively. Various studies indicate different slopes, for instance in \cite{Klypin2001} it was found a slope with $\kappa=1.5$ based on the study of halo simulations, from phase space density arguments, in \cite{TaylorNavarro2001} the density profile was found to show $\kappa=0.75$ instead of $\kappa=1$ and resembles the NFW profile in the outer region, in \cite{Colin2004} based on studies of low mass halos the result was $\kappa=1$ which corresponds to the same limit as NFW, the conclusion in \cite{Diemand2005} is that CDM halos have cusps with $\kappa=1.2$, whereas in \cite{Navarroetal2004} fitting functions with $\kappa=0.7$ were used for radii or the order of $r \sim 0.01$ kpc; the work in \cite{Stoher2006} indicates that within $r \sim 1$kpc the solpe corresponds to $\kappa \sim 1$ and a further extrapolation implies $\kappa \sim 0$ which would correspond to a pseudo-isothermal model. Also in \cite{Navarro2008} non single slope limits are proposed depending on the radial scale considered, for instance for $r \sim 0.1$kpc $\kappa \sim 0.85$, and for $r \sim 1$kpc $\kappa=1.4$..

Observations on the other hand suggest different profiles depending on the type of galaxies considered, for instance dwarf and LSB galaxies are better described by a constant density core model \cite{Burkert1995,Walter2008}, specifically, the mass density profile of dwarf galaxies shows averages of the order $\kappa \sim 0.29$ \cite{OhBlok2010}, whereas LSB galaxies show $\kappa \sim 0.2$ \cite{Blok2001}.   Other analyses including the effects of baryons conclude that $\kappa \sim 0.4$ \cite{Oh2010}. This shows that the problem of the dark matter central distribution in galaxies is still under debate, and there is no clear answer at the moment. Nevertheless, in this paper we look into an even smaller scale of the dark halo, that is, we study the dark matter distribution in the vicinity of a supermassive non-rotating black hole in order to understand whether or not a particular density profile like those mentioned above is favored over the others.

Supermassive black holes in the center of galaxies constitute another brick of the problem, because how they were formed and how they are fed remains a mayor quest. Assuming such black holes exist, it is worth asking their influence in the distribution of matter around them due to their strong gravitational field and determine how they affect in particular the dark matter density profile. As described above, the distribution of dark matter in galaxies is constructed based on simulations obtained from structure formation models, and the limitation of such method is the resolution allowed by currently used algorithms and implementations, which is the reason why the spatial scale of resolution is of the order of kpc. We thus explore how a black hole affects the dark matter around it and determine whether or not the matter distribution is consistent or not with predictions obtained from galactic scale analyses and observations. This approach conversely, faces the problem of being a rather small scale compared to that of dark matter density profiles of galactic scale, however when studying the problem at black hole scale it is possible to infer the behavior of matter in a region within 1pc, whereas the numerical and observational studies above deal with scales of kpc and consider the geometry of the center of the galaxy smooth. However we expect that our results extend up to the kpc scale.

An important feature of studying dark matter around a black hole is that relativistic fluid dynamics equations on a non-flat background space-time are to be solved, unlike the most common approach of supermassive black hole growth and accretion of dark matter models involving a phase space analysis of a Newtonian version of black holes based on the seminal accretion models \cite{LighShap1977,ZhaoRees2002}. Alternatively, in the relativistic regime important bounds on the collisonless dark matter central density have been proposed that allow the observed masses of central black holes \cite{HernandezLee}. In this paper on the other hand, we model the dark matter as a relativistic collisional ideal gas. Even though collisional dark matter has been considered to have astrophysical importance, for instance in non standard models of supermassive black hole growth \cite{Ostriker2000,Hu2006}, as a possible solution to the galactic cusp-core problem \cite{Spergel2000} and the corresponding gravitational collapse \cite{Moore2000}, we also focus on the limit of a collisionless gas, which is the case consistent with the CDM standard approach. 

Two important condition of our model are in turn, one is that we do not consider the effects of baryons and our model consists on the dark matter component only; the second is that we assume the black hole space-time is fixed and the dark matter is a test matter field, which is justified due to the small accretion rate -even in spherical symmetry- happening when the dark matter has pressure,  which is of the order of one solar mass per gigayear. \cite{GuzmanLora}.
We thus solve the fully time-dependent relativistic Euler equations for the fluid on a Schwarzschild black hole space-time, where the test fluid plays the role of dark matter, that is, we implicitly assume the dark matter  behaves as a smooth matter field, unlike the most common approach based on n-body or smoothed particle hydrodynamics simulations. We start up modeling the dark matter as a relativistic ideal gas with pressure, that is, we consider the fluid obeys an ideal gas equation of state, an approach that has been used to estimate upper bounds on accreted mass of collisional dark matter by supermassive black holes  \cite{GuzmanLora}. We have found that there are late-time attractor type of solutions for the fluid density profile with constant accretion rate for the cases of non-zero pressure. Such late-time scenarios can be found starting with rather out of equilibrium initial conditions for the fluid in terms of initial inward velocity, initial rest mass density and adiabatic index. In this paper we analyze the late-time state of the rest mass density profile of such states  in order to envisage a possible consistent dark matter distribution at very local scales near a supermassive black hole.

The paper is organized as follows, in the following section we describe the formulation of Euler's equations we use to model the accretion of dark matter, in section \ref{sec:methods} we describe our numerical methods and in section \ref{sec:results} we present our results.  Finally in section \ref{sec:conclusions} we discuss our results and draw some conclusions.


\section{Relativistic Euler equations}
\label{sec:equations}
 
\begin{figure*}
\includegraphics[width=8cm]{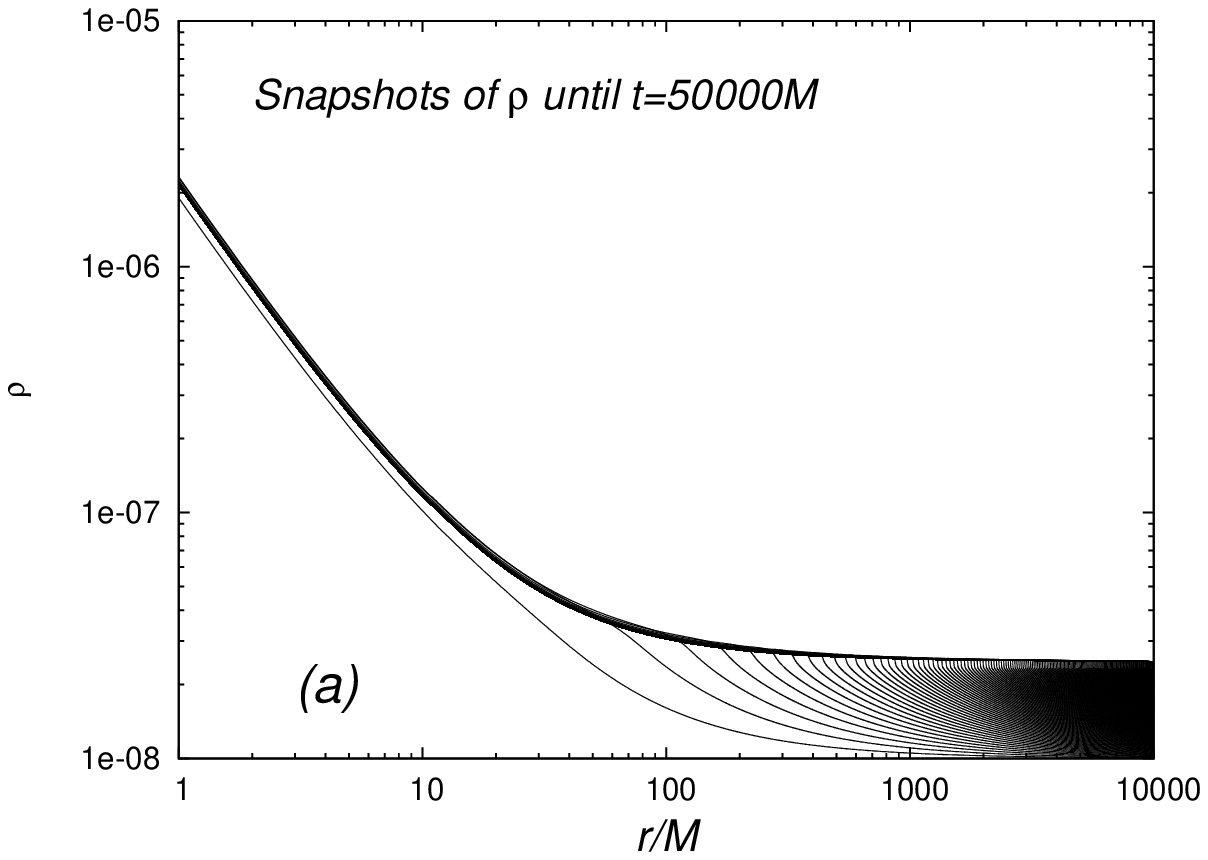}
\includegraphics[width=8cm]{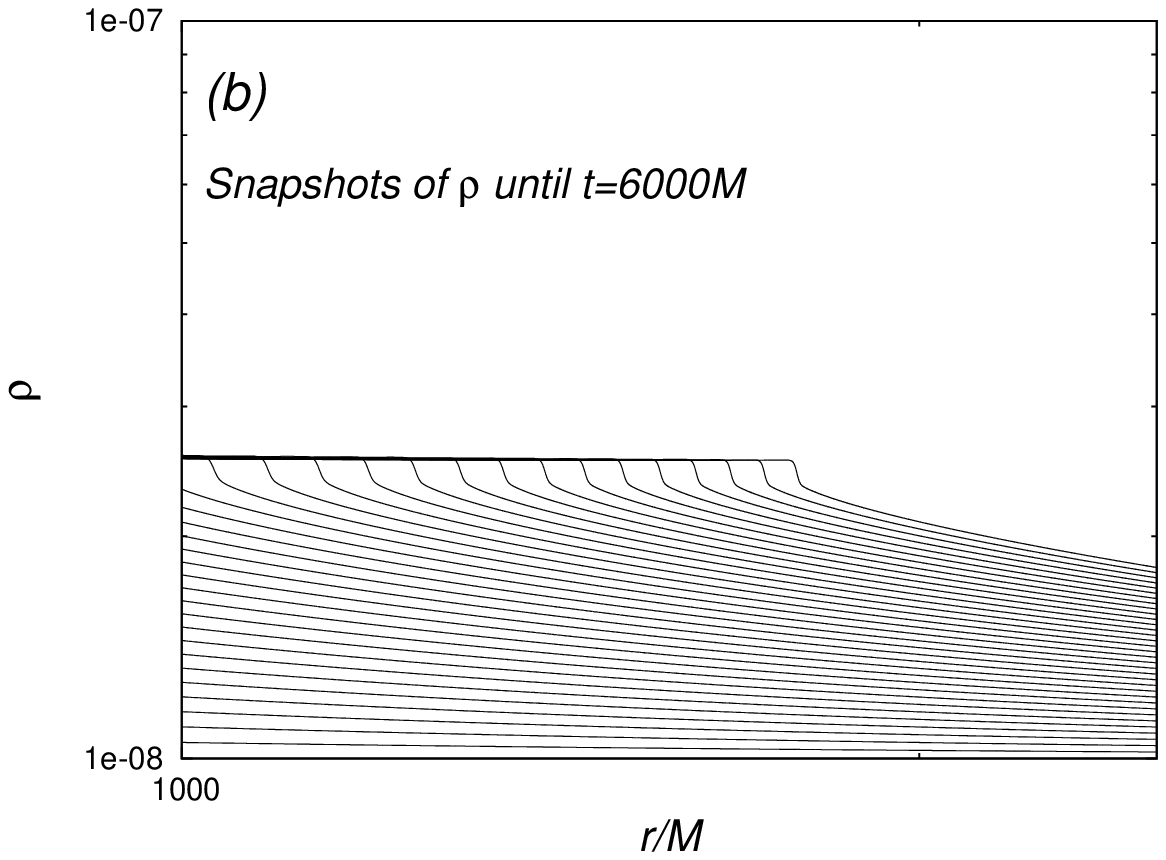}
\includegraphics[width=8cm]{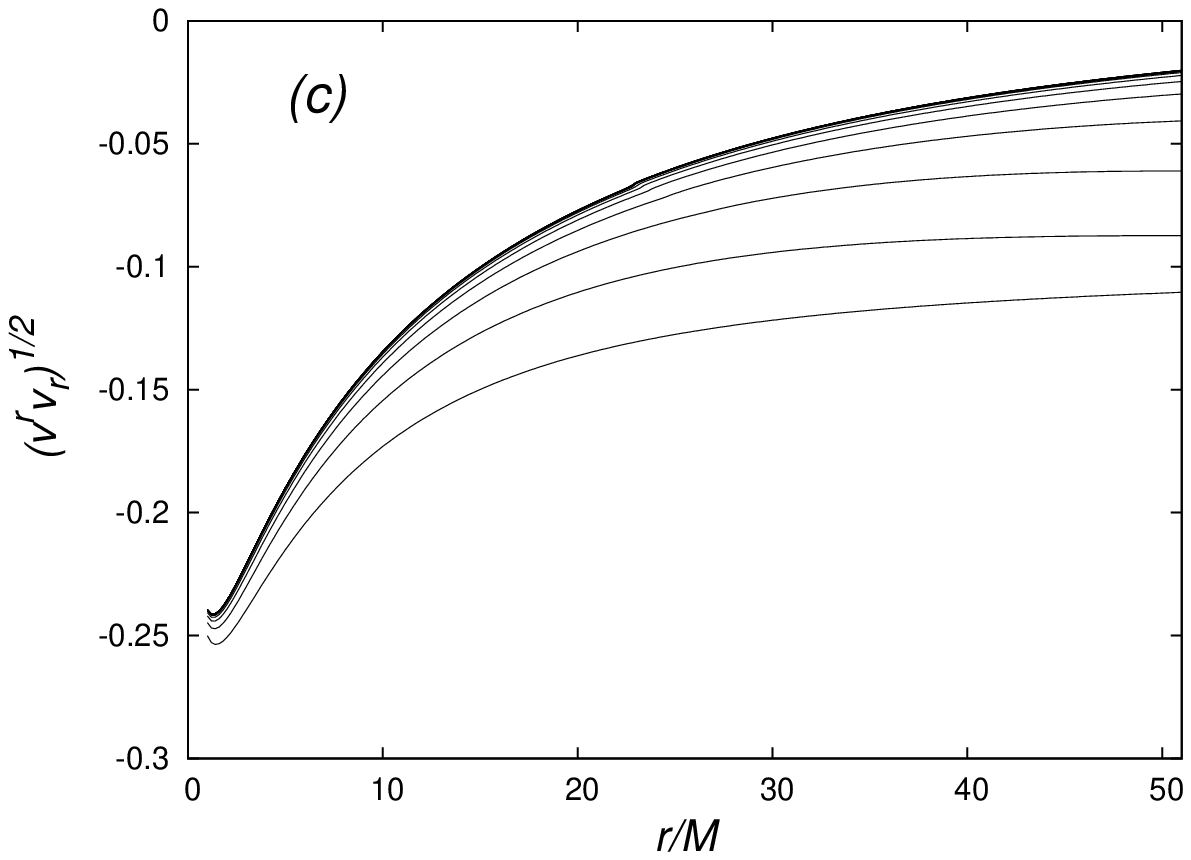}
\includegraphics[width=8cm]{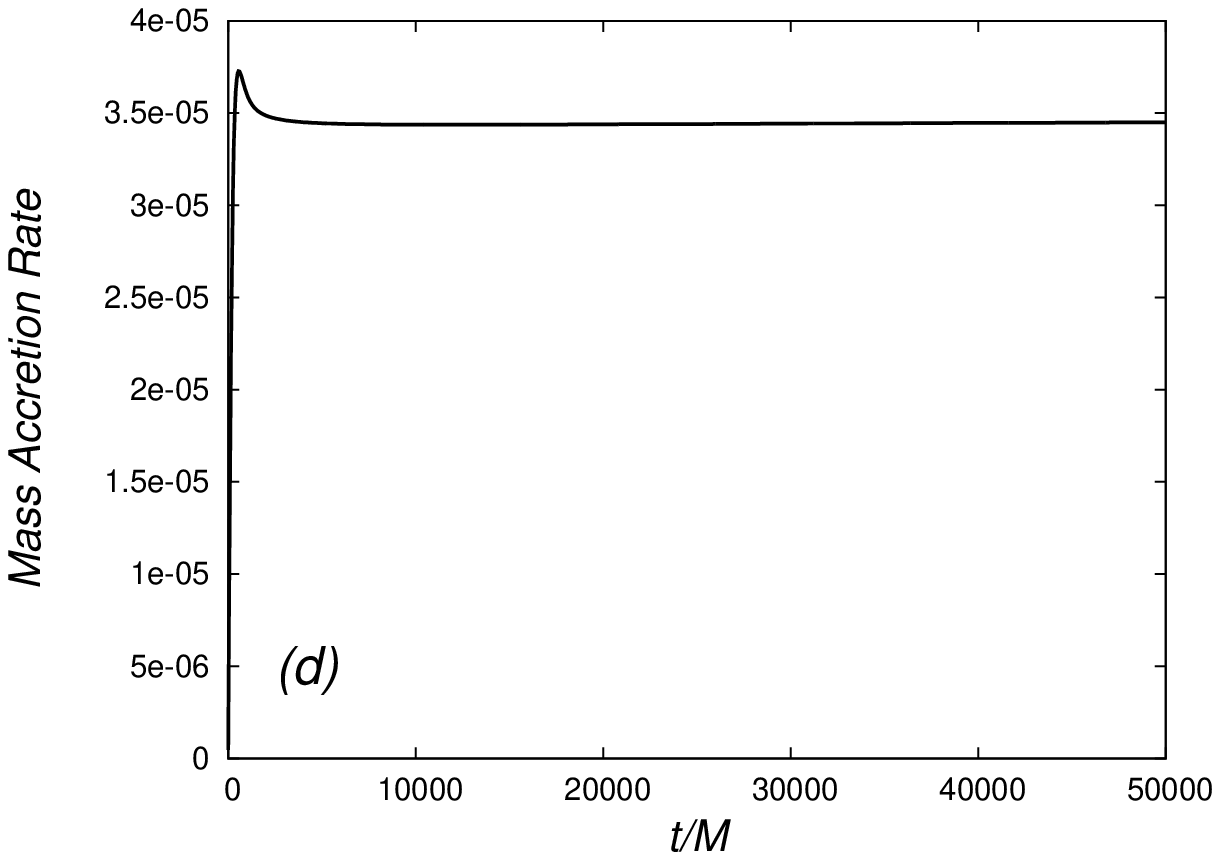}
\caption{\label{fig:late-time} (a) We show snapshots of the density from $t=0$ to $t= 50000M$ every $250M$; earlier times correspond to the lines in the bottom and later times correspond to lines in the upper part approaching the bold black line. The bold black line is actually the superposition of various of the late-time snapshots which shows a nearly stationary state. The density profile shows two clear regimes: one near the black hole up to $r \sim 50M$ which shows a clear polynomial shape and a second one after $r \sim 500M$ which is nearly constant; these are the two domains we use to fit the density profile of the late-time solution. (b) We show a zoom in of a subset of (a) until $t=6000M$ and show that a front shock moving to the right leaves behind the stationary density profile. (c) The velocity of the fluid also reaches a steady state with one quarter the speed of light near the horizon. (d) We also show the accretion mass rate measured at the horizon; this shows how a steady state is reached at late-times. The initial data correspond to $\rho_{0}=10^{-8}$, $v^{r}_{0} = -0.1$, $\Gamma=1.1$ and $\epsilon=0.8$. For the accretion mass rate we use the accretion mass rate formula for the spherical accretion case $\dot{M}_{acc} = -4 \pi r^2  \rho W \left( v^r - \frac{\beta}{\alpha} \right)$.}
\end{figure*}

\begin{figure}
\includegraphics[width=8cm]{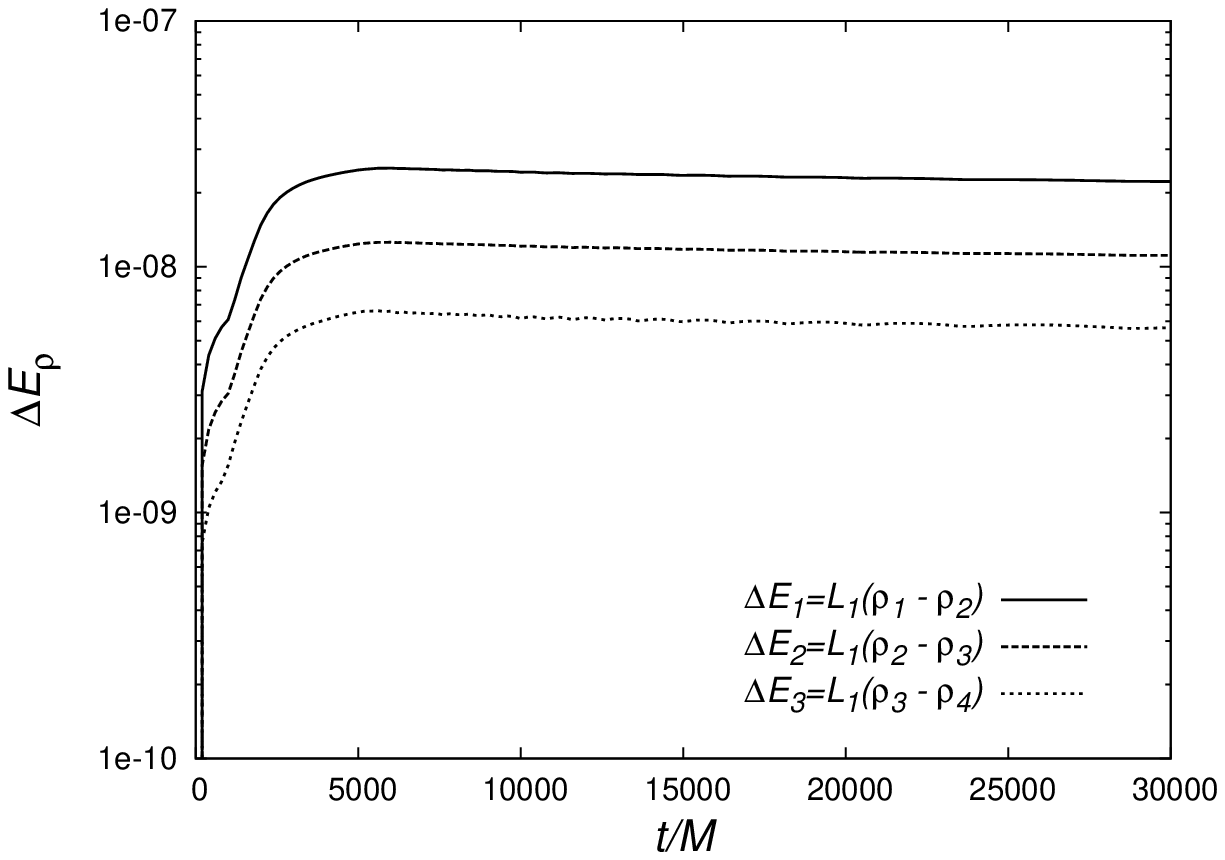}
\caption{\label{fig:convergence} As an illustration of the accuracy of our calculations we show the order of self-convergence of $\rho$, for the case of initial density $\rho_0=10^{-8}$ with $\Gamma=1.06$ and $v^{r}_{0} =-0.1$ for $\epsilon_{0}=0.5$. We calculate the convergence using the $L_1$ norm of the differences between the value of the density for four different resolutions $\Delta x_1=0.4$, $\Delta x_2=\Delta x_1/2$, $\Delta x_3=\Delta x_2/2$ and $\Delta x_4=\Delta x_3/2$, corresponding to the numerical calculation of density $\rho_1,~\rho_2,~\rho_3$ and $\rho_4$ respectively.  Then, the self-convergece factor, $Q$, for these three resolutions is calculated from the following expresion as  $2^{Q_{sc}}=\Delta E_1/\Delta E_2$ where $\Delta E_1 = L_1(\rho_1 - \rho_2)$, $\Delta E_2 = L_1(\rho_2 - \rho_3)$ and $\Delta E_3 = L_1(\rho_3 - \rho_4)$. The dominant error is that of the reconstruction of variables at the interface cells, which is a piece wise constant and therefore first order accurate. We show in this plot that $Q=1$, that is the relationship $\Delta E_i = 2^1 \Delta E_{i+1}$ holds.}
\end{figure}

The system of equations we solve are those of a perfect fluid evolving on a fixed curved background space-time, that is

\begin{eqnarray}
 \nabla_{\nu}(T^{\mu \nu})=0, \label{eq:conserv1}\\
 \nabla_{\mu}(\rho u^{\mu}) = 0, \label{eq:conserv2}
\end{eqnarray}

\noindent where $\rho$ is the proper rest mass density, $u^{\mu}$ is the 4-velocity of the fluid and $\nabla_{\mu}$ is the covariant derivative consistent with the 4-metric of the space-time \cite{mtw}. The stress energy tensor of the fluid in these equations is given by

\begin{equation}
T^{\mu \nu} = \rho h u^{\mu}u^{\nu}+p g^{\mu \nu}. \label{eq:emt}
\end{equation}

\noindent where $p$, $h$ and $g^{\mu \nu}$ are respectively the pressure, the specific enthalpy and the 4-metric tensor of the space-time, respectively. In order to write down explicitly our equations we write down the Schwarzschild metric in the 3+1 fashion for the coordinate system $(t,x^i)$, that is \cite{mtw}:

\begin{equation}
ds^2 = -(\alpha^2 - \beta_i \beta^i)dt^2 + 2\beta_i dx^i dt + \gamma_{ij}dx^i dx^j, \label{eq:lineelement}
\end{equation}

\noindent where $\alpha$ is the lapse function and $\beta^i$ is the shift vector and $\gamma_{ij}$ is the inherited 3-metric on the space-like hypersurfaces. In order to allow the fluid to enter the black hole's horizon we use horizon penetrating coordinates with Eddington-Finkelstein slices, for which 
$\alpha = \frac{1}{\sqrt{1 + \frac{2M}{r}}}$,
$\beta^i=\left [ \frac{2M}{r}\left(\frac{1}{1+\frac{2M}{r}}\right),0,0\right]$ and 
$\gamma_{ij}=diag[1+\frac{2M}{r},r^2,r^2\sin^2\theta]$. For this coordinate choice, equations (\ref{eq:conserv1},\ref{eq:conserv2}) for (\ref{eq:emt}) and (\ref{eq:lineelement}) written in a balance law type based on the Valencia formulation of relativistic hydrodynamics \cite{valencia1991,valencia1997,wai-mo}, which for the spherically symmetric case reads \cite{GuzmanLora}

\begin{equation}
\frac{\partial {\bf u}}{\partial t} + \frac{\partial {\bf F}^r({\bf u})}{\partial r} = {\bf S},\label{eq:FluxConservativeSpherical}
\end{equation}

\noindent where

\begin{eqnarray}
{\bf u} &=& 
	\left[
	\begin{array}{c}
	D\\ 
	J_r \\ 
	\tau 
	\end{array}
	\right]=
	\left [
\begin{array}{c}
\sqrt{\gamma} \rho W \\
\sqrt{\gamma} \rho h W^2 v_r \\
\sqrt{\gamma}(\rho h W^2 -p - \rho W)
\end{array}
\right] , \nonumber\\
{\bf F}^r &=& 
	\left[
	\begin{array}{c}
	\alpha \left(v^r - \frac{\beta^r}{\alpha} \right)D \\
	\alpha\left( v^r - \frac{\beta^r}{\alpha} \right)J_r + \alpha \sqrt{\gamma} p\\
	\alpha \left( v^r - \frac{\beta^r}{\alpha}\right)\tau + \sqrt{\gamma}\alpha v^r p
	\end{array}
	\right],\nonumber\\
{\bf S} &=&
	\left[
	\begin{array}{c}
	0\\
	\alpha \sqrt{\gamma}T^{\mu\nu}g_{\nu\sigma}\Gamma^{\sigma}{}_{\mu r}\\
	\alpha \sqrt{\gamma} (T^{\mu 0}\partial_{\mu} \alpha - \alpha T^{\mu\nu}\Gamma^{0}{}_{\mu\nu})
	\end{array}
	\right].\label{eq:sphericalConservative}
\end{eqnarray}

\noindent In these expressions the 3-velocity measured by an eulerian observer $v^i$ is defined in terms of the spatial part of the 4-velocity of the fluid by $v^i = \frac{u^i}{W} + \frac{\beta^i}{\alpha}$, where $W$ is the Lorentz factor $W=\frac{1}{\sqrt{1-\gamma_{ij} v^i v^j}}$. We close the system of equations  with an equation of state $p=p(\rho,\epsilon)$, where $\epsilon$ is the specific internal energy, defined by $h=1+\epsilon+p/\rho$.
For this we choose the equation of state of a relativistic ideal gas $p = (\Gamma - 1)\rho\epsilon$, where $\Gamma$ is the adiabatic index.

\subsection{ Initial conditions}
 
In order to solve our system of equations as an initial value problem we have to set initial conditions that we choose to correspond to out of equilibrium configurations in order to let the system of equations to drive the fluid (or not) into a relaxed state. Since the nature of dark matter is still unknown we have to be as open as possible and choose out of equilibrium initial conditions with most free parameters as possible. In the first place we choose an initial constant density profile with amplitude $\rho_0 = \rho(t=0)$ which we found to provide quite out of equilibrium initial conditions. Secondly, no initial velocity $v^{r}_{0}$ of the fluid can be assumed, and since we start up with out of equilibrium conditions we choose this parameter to be rather arbitrary. Also, there is no prescription -as far as we can tell- about the internal energy of a gas model for dark matter, thus we use various values of the initial internal energy of the gas $\epsilon_0 = \epsilon(t=0)$. Finally the adiabatic index $\Gamma$ determines how collisional the gas is, and we look for attractor solutions for various values of this parameter as well.


\section{Numerical methods}
\label{sec:methods}
 
We solve our equations on a spatial domain ranging from $M \le r \le 10000M$, that is, from inside the black hole's horizon at $r=M$ where we apply the excision technique \cite{SeidelSuen1992} up to an external boundary located far away from the horizon; the outer boundary in physical units for a black mass of $M=10^9 M_{\odot}$ corresponds to 0.5pc, which is a rather smaller scale compared to previous analyses focused on dark halo profiles based on structure formation; in all our runs we make sure that the external boundary is causally disconnected, so that we avoid the potential influence of boundary conditions applied in such boundary on our results. In order to solve these equations we evolve initial data for the three conservative variables using a finite volume cell centered discretization of the domain, second order accurate High Resolution Shock Capturing method based on the HLLE approximate Riemann solver to calculate the numerical fluxes at the cell interfaces \cite{hlle} and a constant piecewise cell reconstruction of variables which is first order accurate \cite{LeVeque}. The evolution is worked out using the method of lines using a third order Runge-Kutta integrator along the time direction. 
We also implemented a density atmosphere in order to avoid the divergence of our equations such that the rest mass density is always assumed to be $\rho=max(\rho,\rho_{atm})$, where we choose  values between $\rho_{atm}= 10^{-8}$ and $\rho_{atm}= 10^{-12}$ with which we obtain the same physical results and convergence of our implementation, and in fact we also have used such values as the initial density in some of our numerical experiments. Furthermore, we have practiced self-convergence tests that show our implementation converges to first order in the primitive and conservative variables as expected according to our cell reconstruction algorithm.

\section{Results}
\label{sec:results}
 
A pressurless gas system corresponds to $\Gamma=1$, which lacks of spherically symmetric late-time stationary solutions and is the reason why we do not study such system here \cite{GuzmanLora}. Instead, for $\Gamma > 1$ the fluid stabilizes around a situation of constant accretion mass rate and in any case the pressureless case can be studied as a limit when $\Gamma \rightarrow 1$. Therefore, we now focus on this type of configurations and analyze their late-time behavior. As an example of how a late-time solution looks like we show in Fig. \ref{fig:late-time} snapshots of the rest mass density, velocity of the fluid and mass accretion rate for a particular case. In Fig. \ref{fig:convergence} we illustrate the self-convergence of the density for one of our runs, which indicates that our implementation is consistent with the numerical methods and that our results are reliable.

\begin{figure}
\includegraphics[width=8cm]{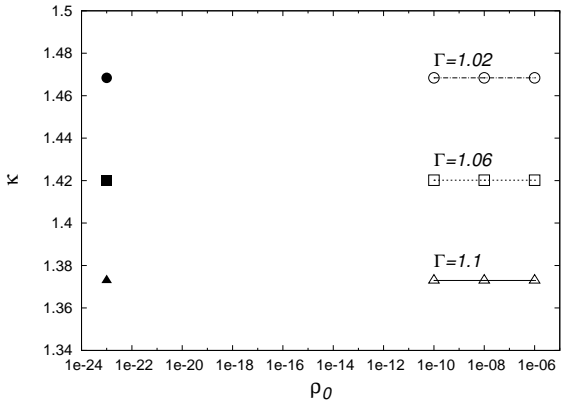}
\caption{\label{fig:alpha_rho_0} We show the values of $\kappa$ for a given $\epsilon_0 =0.2$ and various values $\rho_0$ and the adiabatic index $\Gamma$ for region I and various very different initial conditions. These results show the independence of $\kappa$ on $\rho_0$ for a given combination of $\epsilon_0$ and $\Gamma$. We have made sure that this happens in all our simulations. We also show the extrapolation corresponding to the dark matter density in geometrical units $\rho_0 = 10^{-23}$ which is equivalent to $\rho_0 \sim 100 M_{\odot}/pc^3$, for a black hole mass $M=10^9 M_{\odot}$. The transparent points correspond to the results of our numerical experiments whereas the filled symbols correspond to extrapolations of our results to the dark matter density. Similar results are obtained for region II.}
\end{figure}

We consider different values of the initial rest mass density at initial time $\rho_0$, the adiabatic index $\Gamma$ and the internal energy at initial time $\epsilon_0$.  Once the system relaxes, that is, once it approaches a stationary type of state with constant mass accretion rate and time independent density, pressure and velocity, we fit the resulting density profile with a fitting function of the type $A/r^{\kappa}$, where $\kappa$ is the parameter that determines the slope of the density profile of the late-time attractor type solution. Results shown in Fig. \ref{fig:late-time} suggest two different regions for the fitting of the density profile: region (I) ranging from the event horizon up to $\sim 50M$ where a power law is evident, and region (II), which ranges from approximately $800M$ outwards, which within our domain is of the order of tenths of parsec. Region I is related to the strong field region whereas region II is the one that would eventually match the kpc scale and therefore the results to be compared with the galactic scale results. These regions vary from one simulation to another because the range of the parameter space is wide, however we found $50M$ to be an upper bound of region I in all our cases and the range $800 - 1500M$ the boundary of region II.

Our first result is that for given values of $\Gamma$ and $\epsilon_0$ for different values of $\rho_0$, the parameter $\kappa$ results to be the same as shown in Fig. \ref{fig:alpha_rho_0} for region I. That is, the profile exponent $\kappa$ does not depend on the initial value of the constant density profile $\rho_0$ used to start up our simulations. This allows one to consider only the parameter space to be the $\Gamma - \epsilon_0$ plane and choose a given value of $\rho_0$ that provides better accuracy and use the resulting slope for other values of such initial parameter. In particular we are interested in exploring the behavior of $\rho_0$ that corresponds to a reasonable value of the dark matter density. We illustrate this extrapolation also in Fig. \ref{fig:alpha_rho_0}. We have verified that our stationary solutions reproduce those exact solutions first found by Michel \cite{Michel1972} on the accretion of an ideal gas with pressure. As far as we can tell, no definitive results on the attractor properties of this solution have been discussed before and it is interesting that we have found them as the result of the evolution of rather general initial conditions. There are then two reasons to support our extrapolation: 1) our solutions show a universal behavior for different values of the initial density for various values of  $\Gamma$ and $\epsilon_0$, and 2) Michel's solution would be valid for arbitrary values of the density, however not verified at such ultralow densities as an attractor solution.

\begin{figure}
\includegraphics[width=8cm]{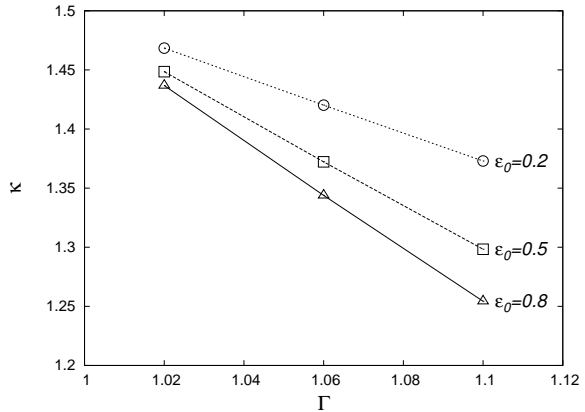}
\caption{\label{fig:alpha} We show the values of $\kappa$ in terms of the adiabatic index and initial internal energy for region I. The exponent strongly depends on the initial conditions. Nevertheless, in the limit of $\Gamma$ approaching one, that is, the collisionless case, the exponent approaches 1.5 independently of $\epsilon_0$.}
\end{figure}

\begin{figure}
\includegraphics[width=8cm]{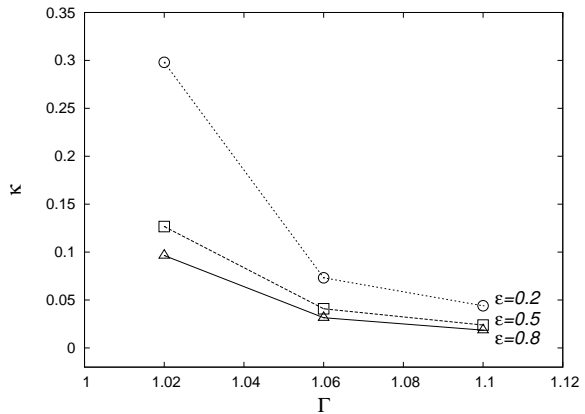}
\caption{\label{fig:alpha2} The values of $\kappa$ in terms of the adiabatic index and initial internal energy are presented. The domain where the density is fitted correspond to $800 \le r/M \le 1500$. As we can see, the bigger the values of $\Gamma$ the smaller the values of $\kappa$.}
\end{figure}

The exploration of the resulting parameter space indicates that $\kappa$ depends monotonically on $\Gamma$ and $\epsilon_0$. These results are summarized in Fig. \ref{fig:alpha} for region I and in Fig. \ref{fig:alpha2} for region II. For each value of $\epsilon_0$ we find different values of $\kappa$ depending on the value of $\Gamma$. If our fluid represents the dark matter, this implies that the preferred slope of the late-time solution for a collisional dark matter candidate depends both, on the initial internal energy and the adiabatic index only. Our results indicate that in order to have a less cuspy profile, that is, smaller $\kappa$, a higher adiabatic index is needed, meaning that the dark matter is more collisional.

Concerning region I, the extrapolation of our results in the limit $\Gamma \rightarrow 1$, that is, the collisionless case, prefers an approximate exponent $\kappa \sim 1.5$ independently of the initial internal energy as seen in Fig. \ref{fig:alpha}. It is also noticed that for bigger values of $\Gamma$ smaller values of $\kappa$ are found.

On the other hand, for region II the results are less coincidental. The first finding is that in the limit $\Gamma \rightarrow 1$ the value of $\kappa$ differs for different initial conditions as shown in Fig. \ref{fig:alpha2}, and no upper bound can be estimated based on our results. Nevertheless, we find that pressure suffices to provide very small values of $\kappa$, in fact smaller than 0.1 for $\Gamma \ge 1.1$, which is a result consistent with flat density profiles related to cored halos.

\section{Conclusions and discussion}
\label{sec:conclusions}

The results in this paper are well far from the most general case, which would involve the non-spherical accretion of matter, on a black hole with non-zero angular momentum and possibly a dynamical geometry ruled by the full non-linear Einstein's equations. The approximation of the fluid to be a test matter field is appropriate in the present case due to the small density values we use and to the quick stabilization of the accretion rate, although the analysis of the rate growth of the event horizon in a full non-linear regime would certainly provide more precise results and would allow one to track the growth of the black hole event horizon mass.

On the other hand, as a first approach, the spherical symmetry is justified and is practical in the sense that our results work as bounds of more general cases involving non-zero angular momentum of matter. At this respect there are studies considering non spherical accretion of matter, for instance \cite{ReadGilmore2003,Holley2006}, all of them based on the phase space analysis focused on the lose-cone refilling ruled by Newtonian gravity and therefore an approximation that deals well with non-trivial impact parameters of the matter but not with the strong gravitational field. Thus a study of matter winds with non-zero angular momentum around black holes would tell us more about not only density profiles of potentially existing attractor in time configurations, but also on the cross section of black holes that would better restrict the accretion rates of dark matter.

Nonetheless, the present first approach results provide some important indications. Under the conditions we used, the density profile is cuspy only in the region near the black hole, whereas the density stabilizes around a nearly constant density profile in the far region. This result implies for instance that for a black hole of mass $10^9 M_{\odot}$, a density profile is nearly flat starting from distances of already $1000M \sim 0.05$ pc on.

Our main conclusion is that if the profile we found in the far region near $\sim 0.1$pc prevails up to scales further than those used here, and this is to be compared to galactic scale analyses based on structure formation approaches and observations, the matter density shows a non-cuspy profile for the parameter space we analyzed here. This means that a nearly constant dark matter density profile is consistent with the presence of a black hole at the price of adding some pressure. Moreover, such profile corresponds to an attractor in time type of solution, that is, a profile that is obtained under a variety of different initial conditions of the matter field.

It is worth to mention here that we are not considering the contribution of baryonic matter, however it is important to emphasize that the effects of baryons on the dark matter distributions plays an important role for instance in the cored shape of dark halos; some of these effects are: feedback \cite{WeinbergKatz2002,Valenzuela2007}, dynamical friction \cite{Zant2001}, triaxial effects \cite{Hayashi2006,Hayashi2007} among others, which should be considered in further relativistic analyses like ours.


\section*{Acknowledgments}

This research is partly supported by grants: 
CIC-UMSNH-4.9 and 
CONACyT 106466.
The runs were carried out in the IFM Cluster.


\bsp

\label{lastpage}

\end{document}